\documentstyle[aps,epsfig]{revtex}
\begin{document}
\draft

\title{Ground state phase diagram of 2D electrons in a high Landau level\\
-- DMRG study --}
\author{Naokazu Shibata and Daijiro Yoshioka}
\address{
Department of Basic Science, University of Tokyo,
Komaba 3-8-1 Meguro-ku, Tokyo 153-8902, Japan
}

\date{january 25, 2001}

\twocolumn[\hsize\textwidth\columnwidth\hsize\csname @twocolumnfalse\endcsname

\maketitle

\begin{abstract}
The ground state phase diagram of 2D electrons in a high Landau level
(index $N=2$) is studied by the density matrix renormalization group 
(DMRG) method. Pair correlation functions are systematically calculated
for various filling factors from $\nu=1/8$ to $1/2$. It is shown that 
the ground state phase diagram consists of three different CDW states 
called stripe-phase, bubble-phase, and Wigner crystal.
The boundary between the stripe and the bubble phases is determined 
to be $\nu^{\mbox{\scriptsize s-b}}_c \sim 0.38$, and that for the 
bubble phase and Wigner crystal is $\nu^{\mbox{\scriptsize b-W}}_c \sim 0.24$. 
Each transition is of first order.
\end{abstract}


\vskip2pc]


The electrons in two-dimensional systems are confined to the 
lowest Landau level under a high perpendicular magnetic field. 
In this limit, Laughlin proposed the ground state many body 
wave function at filling factors $\nu = 1/q$ ($q$ is an odd integer) 
written by the Jastrow-type wave functions\cite{Lagh},
which are exact zero-energy eigenstates of short-ranged repulsive 
interactions. This Laughlin state is an incompressible liquid
with an excitation gap, and the experimental results of 
fractional quantization are explained.

In a weak magnetic fields, however, the higher Landau levels are
occupied by electrons.
The wave function in the higher Landau levels extends over a space
with oscillations, and the electrons are scattered with each other 
over an area with a characteristic $q$ dependence.
This fact grows long range correlations between the electrons, 
and the fractional quantum Hall state becomes unstable.

For high Landau levels with its index $N>1$, 
Koulakov {\it et al.} proposed that the electrons form
charge density waves (CDW's) called stripes and bubbles\cite{Kou1,Kou2}.
The evidence of the CDW's has been experimentally 
observed as anisotropic resistivity and re-entrant integer 
quantum Hall state on ultra high mobility samples\cite{Lill,Du,Coop}.
The formation of the CDW's was recently supported by 
the exact diagonalization studies, and the results
of the exact diagonalizations
are in good agreement with the Hartree-Fock (HF) theory\cite{Rez1,Rez2}.

In spite of such recent development of the theoretical studies,
the detailed properties of the CDW's and the ground 
state phase diagram for high  Landau levels  
are still in question because Koulakov {\it et al.}
used HF approximations, which neglect
the effect of quantum fluctuations, and the exact 
diagonalization studies are restricted to some special filling
and size of systems due to the limitation of available memory
space. Reliable, detailed study is imperative to understand 
the nature of the re-entrant phase and to understand the way
the anisotropy disappears as the filling factor is changed away 
from half-filling.

In this paper we present the numerical results for large 
size systems obtained by the density matrix renormalization group 
(DMRG) method \cite{DMRG} which is applied to the 2D electron 
systems in a high Landau level of $N=2$.
The calculated pair correlation functions show 
that the ground state phase diagram consists of 
three CDW states, the stripe phase, bubble phase 
with only two-electron bubbles, 
and Wigner crystal. The obtained phase diagram is similar to 
that of the HF calculations except that there is no bubbles 
with more than two electrons per bubble\cite{HFn}. 
The boundary between the stripe phase and the 
bubble phase is shown to be $\nu^{\mbox{\scriptsize s-b}}_c \sim 0.38$, 
and that for the bubble phase and Wigner crystal is 
$\nu^{\mbox{\scriptsize b-W}}_c \sim 0.24$. It is also clarified that each 
transition is of first order.

To deal with large size systems, we use the DMRG algorithm\cite{DMRG},
which was originally developed for 1D quantum systems. 
In this method we can calculate the ground state wave function 
and the energy with high accuracy. 
The outline of the algorithm is summarized as follows:
We start from small finite systems, ie. four-site system, and 
divide the system into two blocks. Then add new sites at the end of 
two blocks and expand the blocks with restricting the number of basis states.
The restriction of the basis states is carried out
by keeping only eigenstates of large eigenvalues of the
density matrix which is calculated from the ground state wave function.
Thus the numerical error due to the truncation of basis states is 
estimated from the eigenvalues of the density matrix which are truncated off,
and the accuracy of the wave function is systematically improved
by increasing number of states kept in the blocks.
We repeat the expansion of the blocks, and finally get desired 
size of system within a controlled accuracy.

Since the above algorithm is designed for 1D systems,
we have to find appropriate mapping to a 1D model. 
In this study we use the eigenstate of free electrons 
as local basis, and represent the wave function in the Landau gauge. 
Since each wave function in Landau 
gauge is uniquely identified by the $x$-component of the center 
coordinates $X_j=2\pi\ell^2 j /L_y$\ ($j$\ :\ integer),
we can map the present 2D system onto a 1D lattice model.

The important difference between the present model and usual 
1D quantum systems is that the present model has 
additional conserved quantity, the center of mass of electrons, 
$\sum_{i=1}^{N_e} X_i$. This is due to the conservation of $y$-momentum
$\sum_{i=1}^{N_e} p^y_i$ in the original two dimensional system,
where $p^y_i$ is related with $X_i$ as $p^y_i=X_i/\ell^2$ 
under the Landau gauge.
This conservation law causes some technical problems
in the infinite system algorithm of the DMRG. To avoid
this problem we have to keep additional basis states 
which are not included in the density matrix of the ground state.
However, after we switch to the finite system algorithm, 
we need not care such problems.

The Hamiltonian for electrons in Landau levels
contains only the Coulomb interactions. After the projection 
onto the $N$th Landau level, the Coulomb interaction is written as
\begin{equation}
H=\sum_{i<j} \sum_{\bf q} e^{-q^2/2} \left[ L_{N}(q^2/2) \right] ^2 V(q) 
e^{i{\bf q} \cdot ({\bf R}_i-{\bf R}_j)} ,
\label{Coulomb}
\end{equation}
where ${\bf R}_i$ is the guiding center coordinate of the
$i$th electron, $L_{N}(x)$ are the Laguerre polynomials,
and $V(q) =2\pi e^2/q$ is the Fourier transform of the
Coulomb interaction. The magnetic length $\ell$ is set to be 1.
We consider uniform positive background charge to cancel the
component at $q=0$ in Eq.~\ref{Coulomb}, and neglect the 
electrons in fully occupied lower Landau levels.
In the following we calculate ground state wave function of the 
Hamiltonian for high Landau level of $N=2$
using both the infinite system and finite system algorithms 
of the DMRG. We study various size of systems with up to 18 electrons
in the unit cell of $L_x\times L_y$ with periodic boundary 
conditions in both $x$ and $y$ 
directions. The truncation error in the DMRG calculation is
typically $10^{-5}$ for 18 electrons with 200 states in each blocks.
The existing results of exact diagonalizations are 
completely reproduced within the truncation error.
Since the present Hamiltonian has the particle-hole symmetry, 
we only consider the case of $\nu \le 1/2$\cite{Com1}.

In Fig.~1 we show the ground state pair correlation functions 
in guiding center coordinates for various $\nu$.
The guiding center correlation functions are obtained 
by omitting Laguerre polynomials in the Fourier transformation,
that means we leave out Hermite polynomials from the single
electron wave function when we calculate correlation functions.
It is essentially the correlation functions of
the center of the cyclotron motion, and is identical to
the usual correlation functions for electrons in 
the lowest Landau level.

In Fig.~1 we find clear discontinuity in the 
shape of the correlation function
between $\nu=2/5$ and $4/11$, and between $\nu=1/4$ and $2/9$.
Since we have set the $x$ and $y$ axes to give continuous 
change as much as possible, the discontinuity in Fig.~1 means 
drastic change in the pattern of the correlation function.
Thus the transition 

\begin{figure}[t]
\epsfxsize=75mm \epsffile{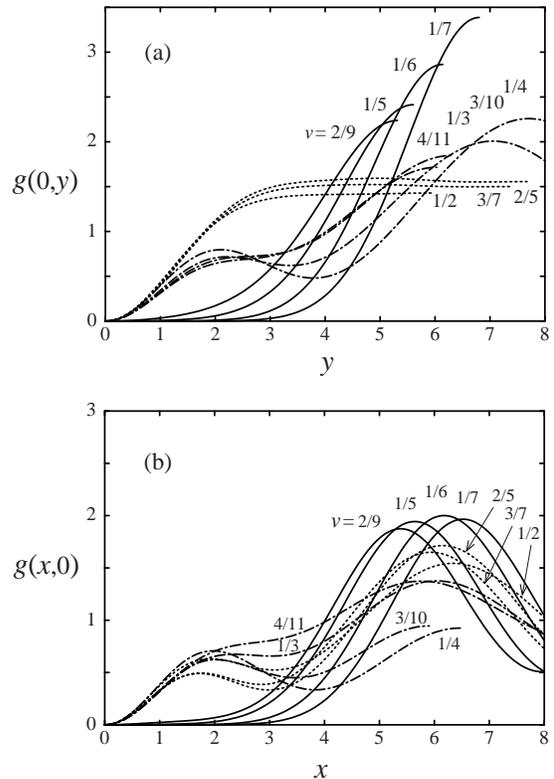}
\caption{
Ground state pair correlation functions in the guiding center coordinates
for electrons in high Landau level of N=2. The magnetic length $\ell$ is
set to be unity.
The number of electrons in the unit cell is 18 for $\nu=1/2$ and
8 for $\nu=1/7$. The aspect ratio $L_x/L_y$ is chosen to obtain 
maximum energy gain around $L_x/L_y=1$.
}
\end{figure}

\begin{figure}[t]
\epsfxsize=85mm \epsffile{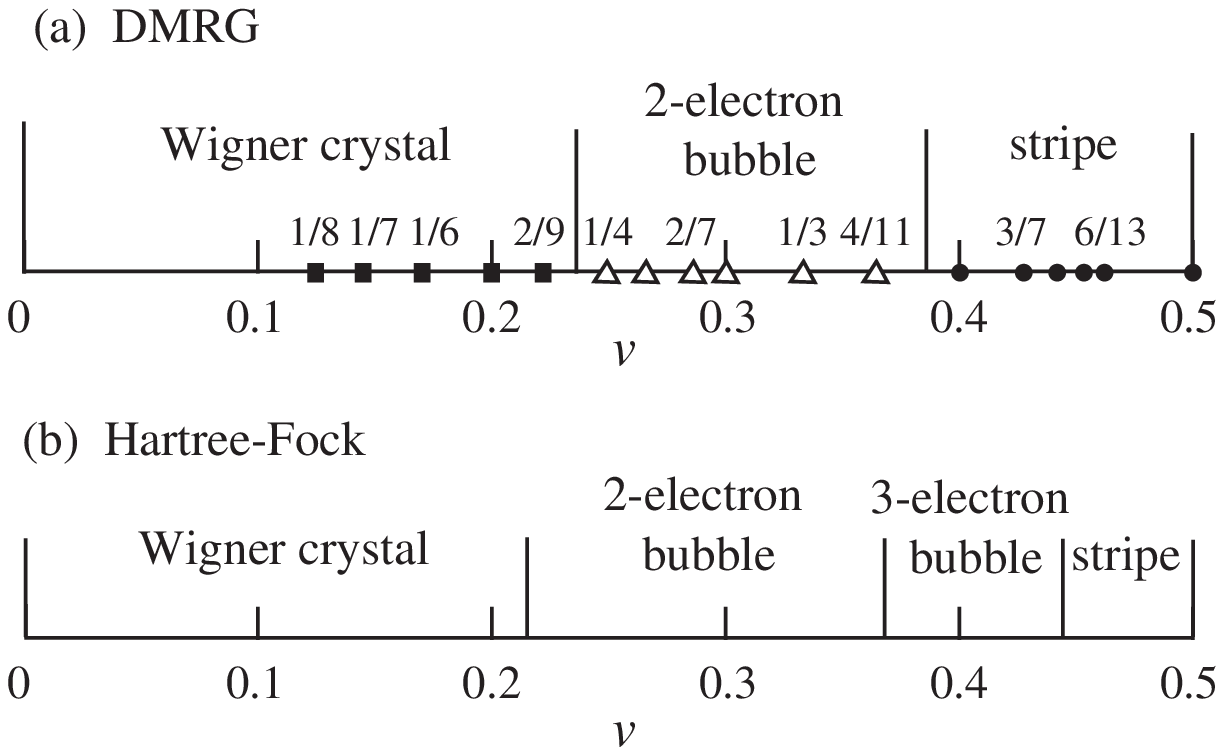}
\caption{
Ground state phase diagram of 2D electrons in high Landau level of N=2
obtained by (a) the DMRG and (b) the Hartree-Fock theory[10].
}
\end{figure}

\noindent
is first order. As shown bellow, the pattern of the correlation function
is characterized by stripes between $\nu=1/2$ and $2/5$, and
bubbles between $\nu=4/11$ and $1/4$, and Wigner crystal below $\nu=2/9$.
Hence, we obtain the phase diagram shown in Fig.~2 (a).
In the following we 

\begin{figure}[t]
\epsfxsize=85mm \epsffile{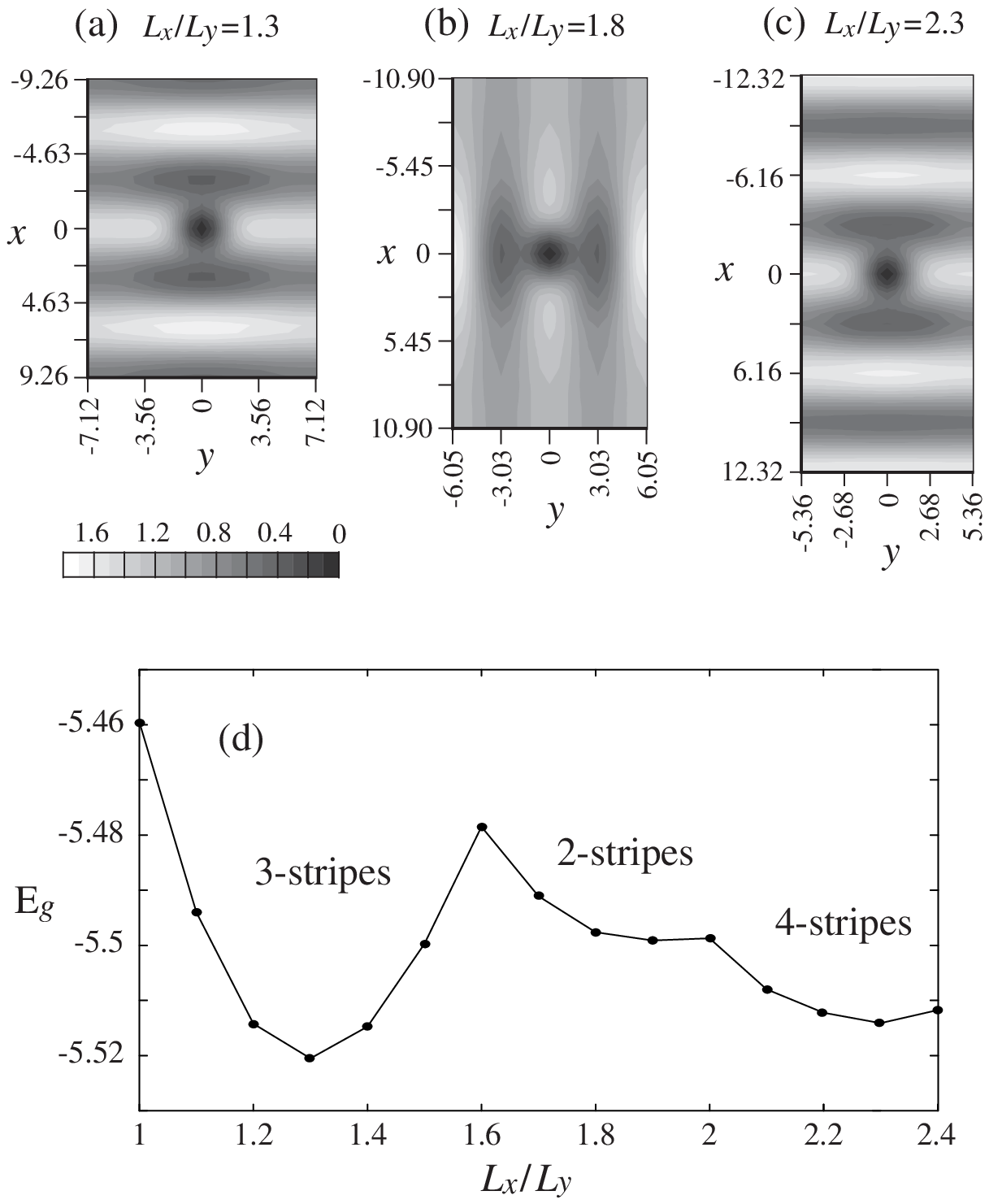}
\caption{
Ground state pair correlation functions in the guiding 
center coordinates and the energy at $\nu=3/7$. 
 The number of electron is 18.
(a) Pair correlation functions for $L_x/L_y=1.3$, (b) 
$L_x/L_y=1.8$, (c) $L_x/L_y=2.3$. (d) The ground state 
energy.}
\end{figure}

\noindent
show detailed structure 
of the correlation functions in each phase.

We start from the stripe phase which appears around the half filling.
In Fig.~3 we show the pair correlation functions at $\nu=3/7$
in the guiding center coordinates.
In this figure, we clearly observe the stripe structure. 
The similar stripe structure is obtained in the HF calculations,
but we find no clear modulations that is predicted 
in the HF theory\cite{Com2}. 
The clear stripe structure similar to the present result 
is obtained also for $\nu=1/2, 6/13, 5/11, 4/9$, and $2/5$ 
with different number of electrons and size of systems 
as expected from Fig.~1.

The detailed structure of the stripes such as the mean separation 
depends on the aspect ratio $L_x/L_y$. To determine the optimal 
stripe structure, we next compare the 
ground state energy.
In Fig.~3 (d) we show the ground state energy for various aspect ratio.
In this figure we find a minimum at $L_x/L_y=1.3$. 
At this ratio the mean separation of the stripes is 6.2,
which is close to the results 6.0 obtained by the HF theory. 
With increasing $L_x/L_y$, both the mean separation and
the ground state energy increases. 
At $L_x/L_y=1.6$, the orientation and the number of the stripes 
in the unit cell are changed due to the level crossing of the
ground state. Then both the mean separation and  ground 
state energy decrease.
The energy takes minimum again around $L_x/L_y=1.9$, where 
the mean separation is 5.9. 
Further increasing $L_x/L_y$,

\begin{figure}[b]
\epsfxsize=75mm \epsffile{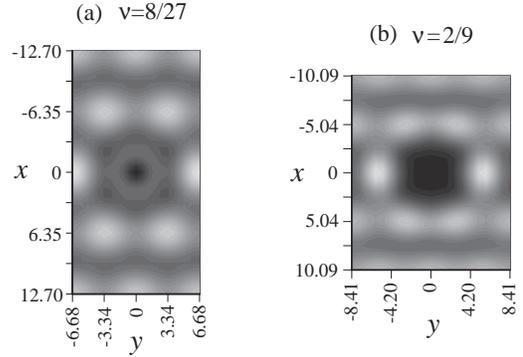}
\caption{
Pair correlation functions in the guiding center coordinates at 
(a) $\nu=8/27$ with 16 electrons, (b) $\nu=2/9$ with 12 electrons.
}
\end{figure}

\noindent
level crossing occurs again and
the number of stripes increases to four in the unit cell. 
Then the ground state energy takes minimum at $L_x/L_y=2.3$,
where the mean separation is 6.2. 
Thus the optimal structure where the energy takes minimum does not 
depend so much on $L_x/L_y$. Since the interval between the minima
that appears on the $L_x/L_y$-axis becomes shorter with increasing 
the size of the unit cell, the optimal structure at the energy 
minimum will be realized in the bulk limit for any $L_x/L_y$.

Now we switch to the bubble phase. The correlation 
function in the guiding center coordinates at $\nu=8/27$ 
for 16 electrons is shown in Fig.~4 (a). 
The aspect ratio is chosen to be 
$1.9$ where the minimum energy is obtained.
In this figure we find 8 bubbles in the unit cell on the triangular
lattice. Since the number of electrons is 16 in the unit cell,
two electrons are clustering together in the guiding center coordinates. 
This pairing of the two electrons makes ring structure in the 
correlation functions around the origin.
As expected from Fig.~1, the same pattern of the two-electron bubbles
is obtained also for $\nu=4/11,1/3,4/13,3/10,2/7,4/15$, and $1/4$, 
and the lattice spacing increases with decreasing $\nu$. 
The similar two-electron bubbles are obtained in the HF calculations.
However, as shown in Fig.~2 (b), the HF theory predicts also the 
three-electron bubbles, each of which contains three electrons.
Since we cannot find three-electron bubbles in the present study,
we think the energy gain due to the quantum fluctuations is relatively 
small for three-electron bubbles.

In the usual electron coordinates the pair correlation functions 
are almost circularly symmetric around the origin as shown in 
Figs.~5 (a) and 6. 
This symmetric correlations contrast to the anisotropic
correlations in the stripe phase.
In Fig.~6, we also plot the result at $\nu=2/9$ where the
ground state is the Wigner crystal. 
We find the enhancement over the case of $\nu=2/9$ around $r \sim 2.5$.
This is caused by the clustering of the electrons in the bubble phase.
Similarly to the guiding center correlations, 
this enhancement makes a ring structure around the origin 
as shown in Fig~5 (a).

\begin{figure}[t]
\epsfxsize=85mm \epsffile{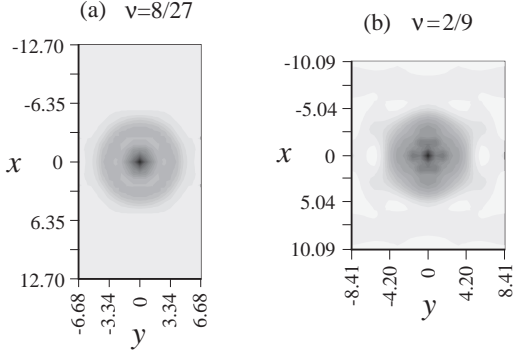}
\caption{
Pair correlation functions in the electron coordinates. 
(a) $\nu=8/27$, 16 electrons. (b) $\nu=2/9$, 12 electrons. 
}
\end{figure}

\begin{figure}[t]
\epsfxsize=75mm \epsffile{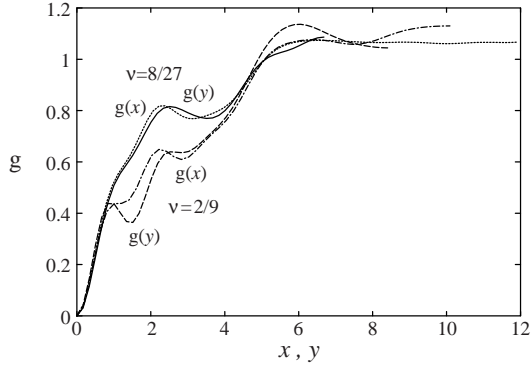}
\caption{
Pair correlation functions in the electron coordinates
for $\nu=8/27$ with 16 electrons, and $\nu=2/9$ with 12 electrons.
}
\end{figure}

Finally, we consider the ground state at low density 
$\nu \stackrel{<}{_\sim} 2/9$.
In the limit of $\nu \rightarrow 0$ the electrons are
separated from each other. 
When the distance to the other electrons exceeds 
the typical length of the single electron wave function,
we expect the difference of the Landau levels becomes 
almost negligible. Thus we expect the formation of Wigner 
crystal as in the lowest Landau level. 
The result for $\nu=2/9$ with 12 electrons shown in Fig.~4 (b)
actually shows that 
the center of the cyclotron motion forms triangular lattice
with 12 lattice points in the unit cell.
This shows the tendency to form Wigner crystal.
Thus we expect the ground state is the Wigner crystal for 
$\nu \stackrel{<}{_\sim} 2/9$.
Even in the electron coordinates shown in Figs.~5 (b) and 6,
the correlation function has peaks at the triangular lattice 
points and the hexagonal symmetry is clearly seen. 
We expect clear crystallization for smaller $\nu$.

Thus we have obtained a reliable phase diagram for a system in the third
lowest Landau level ($N$=2) in the strong magnetic field.
Since we have neglected the spread of the wave function in the third
dimension, and screening effect by electrons in the lower Landau levels, 
the phase boundary may have slightly different value in the actual system.
The absence of the three-electron bubble phase is consistent with the
experiment. This phase is predicted by the HF theory\cite{HFn},
and has not been denied by the exact diagonalization study\cite{Rez2}.
From the phase diagram we can speculate that the re-entrant phase is
the two-electron bubble phase. The coexistence of the Wigner crystal
and the bubbles at the phase boundary around $\nu=1/4$ brings finite 
dissipation into the system separating the two integer quantum Hall 
states. This idea would be jeopardized if the three-electron bubble 
were realized, since then there would be another re-entrant phase.

Part of numerical calculation is performed in the ISSP,
University of Tokyo. The present work is supported by
Grant-in-Aid No.~12640308 and No.~11740184 from Ministry 
of Education, Science, Sports and Culture of Japan.


\begin{thebibliography}{99}
\bibitem{HFn}The details of the Hartree-Fock calculation will be published 
elsewhere.
\bibitem{Com1}The filling factor in this paper is that of the relevant 
Landau level.
\bibitem{Com2}The lack of the modulation is related to the fact that the
transition to the bubble phase is of first order.
\end{thebibliography}
\end{document}